\documentclass[aps,prl,twocolumn,showpacs,superscriptaddress,10pt]{revtex4-1}
\usepackage[colorlinks ,linkcolor=blue,anchorcolor=blue,citecolor=blue,urlcolor=blue]{hyperref}
\usepackage{amsmath,amssymb}
\usepackage{bm}
\usepackage{graphicx}
\usepackage{epstopdf}
\usepackage{color}

\usepackage{float}
\usepackage{color}

\usepackage{tikz}
\newcommand*{\circled}[1]{\lower.7ex\hbox{\tikz\draw (0pt, 0pt)%
    circle (.5em) node {\makebox[1em][c]{\small #1}};}}
\usepackage[T1]{fontenc}
\usepackage{tgtermes}
\begin{document}

\title{Coexisting N\'eel and charge density wave orders in attractive three-color fermions}
\author{Xiang Li}
\affiliation{School of Physics and Technology, Wuhan University, Wuhan
430072, China}

\author{Yu Wang}
\email{yu.wang@whu.edu.cn}
\affiliation{School of Physics and Technology, Wuhan University, Wuhan
430072, China}

\begin{abstract}
   In optical lattices attractive ultracold fermions with three hyperfine-spin components (colors) can form three fermionic configurations depending on interactions: unbound fermion, on-site trion and off-site trion, leading to the coexistence of multiple Fermi species in the ordered phase, which manifests that the attractive three-color fermions are unique from other correlated fermion systems and may host intriguing phases and phase transitions.  At temperature below the superexchange energy scale, we employ the determinant quantum Monte Carlo (QMC) method to investigate the phases and phase transitions in the half-filled attractive three-color Hubbard model on a honeycomb lattice where Hubbard interactions are color-dependent (anisotropic interactions) and the coupling between color 3 and colors (1, 2) serves as a control parameter. In the coupling regime where on-site and off-site trions coexist, our QMC simulations demonstrate coexisting N\'eel and charge density wave orders which are common in condensed matter but rare in ultracold atoms. At weak coupling where the color superfluid (CSF) order is scattered by color-3 fermions, we find that very small coupling of color 3 with colors (1, 2) can destroy the CSF order and the vanishing of the CSF order is not immediately accompanied by the emergence of the on-site trionic phase, which strikingly disagrees with the prevalent results of dynamical mean-field theory. The underlying mechanisms of the coexisting charge/spin orders and the CSF order breaking are also presented based on intuitive physical pictures.

\end{abstract}

\maketitle
\textit{Introduction}.---
SU(3) physics with attractive three-color (hyperfine-spin component) ultracold fermions such as $^{6}\mathrm{Li}$ and $^{40}\mathrm{K}$ atoms \cite{Ottenstein2008,Huckans2009,Regal2003}, is one of major research foci at the interdisciplinary frontiers of ultracold atom physics, condensed matter physics and high energy physics. The attractive SU(3) Fermi gas with tunable interaction is an ideal system for studying trimer states which is hard to be realized in spin-1/2 Fermi gas due to the Pauli exclusion \cite{Mattis1984,Rudin1985,Mattis1986}. In optical lattices loaded with attractive SU(3) ultracold fermions, the variational \cite{Rapp2007,Rapp2008}, self-energy functional \cite{Inaba2009,inaba2011color}, dynamical mean-field theory (DMFT) \cite{Titvinidze2011,Koga2017}, and Bethe ansatz \cite{chetcuti2023probe} studies have demonstrated the phase transition between the color superfluid (CSF) phase and the on-site trionic phase, which is strongly reminiscent of the phase transition between the quark superfluid state and the baryon state in high energy physics \cite{Fodor2002,Aoki2006,Wilczek2007}. Moreover, DMFT studies \cite{Okanami2014,Miyatake2011} have also shown a direct phase transition between the CSF phase and the on-site trionic phase in the attractive three-color Hubbard model where attractions are color-dependent thus breaking SU(3) symmetry of the Hamiltonian. Besides the on-site trion, more interestingly, the exact diagonalization study of the three-fermion attractive SU(3) Hubbard model on the honeycomb lattice suggests another type of three-body bound state, known as the off-site trion, composing of two fermions at one site and one fermion at the nearest-neighbour site \cite{Pohlmann2013}.

In recent years, the long-ignored role of off-site trions in many-body physics of attractive three-color fermions has been attended. In an one-dimensional lattice away from half filling, the density matrix renormalization group (DMRG) studies have shown that off-site trions can develop quasi-long-range correlations in the attractive SU(3) Hubbard model with suppressed on-site triple occupancy \cite{Kantian2009} as well as in the attractive three-color Hubbard model with significantly anisotropic interactions \cite{Capponi2009}. Recently, quantum Monte Carlo (QMC) simulations of the half-filled attractive SU(3) Hubbard model on a honeycomb lattice manifest that on-site trions (majority) and off-site trions (minority) coexist in the charge density wave (CDW) state \cite{xu2019,li2021}, and the off-site trion arising from density fluctuations forms a local bond state \cite{xu2019}. Unlike conventional transitions that only a single Fermi species gets involved in, off-site trions though small in number, together with on-site trions participate in the CDW ordering in the attractive SU(3) Dirac fermions, and particularly the quantum critical point affected by multiple Fermi species can not be described using the conventional Gross-Neveu-Yukawa paradigm \cite{xu2019}.

In this letter, we explore quantum phase transitions in the half-filled attractive three-color Hubbard model on a honeycomb lattice. The coupling between the third color and the other two colors serves as a control parameter of phase transitions. On one hand, breaking SU(3) symmetry results in "magnetic" off-site trions (two ends of which may carry net colors) and on the other hand, reducing the coupling of the third color with the other two colors (namely, introducing "interaction anisotropy") enhances the density fluctuations, which in turn lead to a rise in the number of off-site trions. Thus long-range magnetic correlation of off-site trions may develop in the attractive three-color Hubbard model. Our sign-problem-free QMC simulations demonstrate coexisting N\'eel and CDW orders via tuning the "interaction anisotropy" in attractive three-color fermionic atoms, while previously coexistence of charge and magnetic orders usually occurs in condensed matter but is rare in ultracold atoms \cite{CuZhang2002,CuKivelson2003,ironBalatsky2010,ironDai2012,multiferroicsBayaraa2020,multiferroicsZhang2020,
multiferroicsSolovyev2021,multiferroicsZhang2022,CrSinger2016,CrHu2022,FeQian2001,FeKubetzka2005,FeMeckler2009,FeNiklas2013,
FeHsu2016,InSeStepanov2022,rareGalli2002,rareHanasaki2017,rareSalamatin2018}. In addition, our QMC results disagree with the DMFT prediction that a direct phase transition between the CSF order and the on-site trionic order may occur in the half-filled attractive three-color Hubbard model.

\label{sec:model}
\textit{Model and method}.--- The half-filled attractive three-color Hubbard model on the honeycomb lattice is written in the form
\begin{equation} \label{main.Eq.1}
H=-t\sum_{\langle ij\rangle,\alpha}(c^{\dagger}_{i\alpha}c_{j\alpha}+h.c.)+\sum_{i,\alpha<\beta}U_{\alpha\beta}(n_{i\alpha}-\frac{1}{2})(n_{i\beta}-\frac{1}{2}),
\end{equation}
where $\langle ij\rangle$ denotes a pair of nearest-neighbor sites; $\alpha$ and $\beta$ are the color indices taking only values  $1$, $2$, $3$; $t$ is the nearest-neighbor hopping integral; $n_{i\alpha}=c^{\dagger}_{i\alpha}c_{i\alpha}$ is the particle number operator of color $\alpha$ defined on site $i$; $U_{\alpha \beta}(<0)$ is the attractive interaction between fermions carrying different colors. The chemical potential vanishes at half filling. Inspired by the theoretical and experimental works regarding the SU($N$)-symmetry breaking interactions \cite{Cazalilla_2009,Cazalilla_2014,Huang2020,Huang_2022}, we set $U_{12}=U$ and $U_{13}=U_{23}=U'$ throughout the paper. Note that $|U'|/|U|$ characterizes the interaction anisotropy. When $|U|=|U'|$, the interaction is color-independent (isotropic) and the Hamiltonian Eq.~(\ref{main.Eq.1}) possesses SU($3$) symmetry; when $0<|U'|<|U|$, the interaction anisotropy is introduced and the SU(3) symmetry of the Hamiltonian Eq.~(\ref{main.Eq.1}) is reduced to SU($2$)$\otimes$U($1$); when $|U'|=0$, the system is decoupled into two subsystems, namely SU(2) interacting fermions and free fermions, owning SO($4$)$\otimes$U($1$) symmetry \cite{Yang1990}.

The determinant QMC simulation of the half-filled attractive three-color Hubbard model is sign-problem-free  when the Hubbard-Stratonovich decomposition in the spin-flip channel is employed \cite{Wang2015,xu2019,li2021}. We employ the determinant QMC method at $T=0.1t$ (below the superexchange energy scale) to simulate the phase transitions in the attractive three-color Hubbard model. The simulations are performed on honeycomb lattices with sizes $L=3,6,9,12$ at half filling. Unless specifically stated, the color-dependent Hubbard interaction $U_{\alpha\beta}$ are given in the unit of $t$.

\label{mag}
\textit{Coexisting spin/charge orders}.--- To gain some insight into the physics of half-filled attractive three-color Hubbard model, we first consider the two-site model. For infinite coupling, an on-site trion forms. In the strong-coupling limit, the perturbation theory gives the first-order correction term which is cast in the form of an off-site trion
\begin{equation}
\label{off trion}
\begin{aligned}
|\psi_{\text{off}}\rangle=&-\frac{t}{|U|+|U'|}\left(|12,3\rangle+|13,2\rangle+|23,1\rangle \right)\\
&+\left(-\frac{t}{2|U'|}+\frac{t}{|U|+|U'|} \right)|12,3\rangle.
\end{aligned}
\end{equation}
We see that the off-site trion takes a N\'eel-like configuration due to the second term on the right hand side of Eq.~(\ref{off trion}) (The first term is SU(3) symmetric and "colorless"), when interactions become anisotropic, $|U'|<|U|$.

In the trionic CDW regime where on-site and off-site trions coexist, density fluctuations lead to the formation of off-site trions from on-site trions \cite{xu2019,li2021}. Interaction anisotropy $|U^\prime| < |U|$ enhances density fluctuations, thus driving up the number of off-site trions. Our determinant QMC simulations will demonstrate that interaction anisotropy can induce the N\'eel ordering of correlated off-site trions on the background of trionic CDW phase. In our simulations, we set $|U|=6$ in order to broaden the coupling range of $|U^\prime$| ($<|U|$) in which on-site and off-site trions coexist \cite{xu2019}.

\begin{figure}[t]
  \centering
  \includegraphics[width=0.75\linewidth]{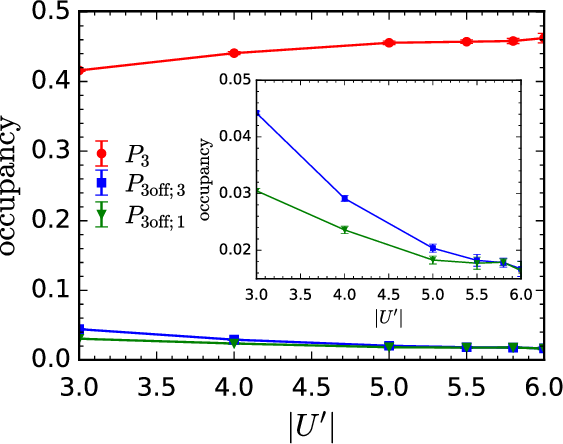}
  \caption{The on-site triple occupancy $P_{3}$, off-site triple occupancies $P_{3\mathrm{off};1}$ and $P_{3\mathrm{off};3}$ are plotted as a function of $|U'|$. The inset is a
   zoom-in view of $P_{3\mathrm{off};1}$ and $P_{3\mathrm{off};3}$ curves. The lattice size is $L=9$.}
  \label{occupation}
\end{figure}

The structures of trions can be described, respectively, by the on-site triple occupancy
\begin{equation}
P_{3}=\frac{1}{N}\sum_{i}\langle n_{i1}n_{i2}n_{i3}\rangle
\end{equation}
and the off-site triple occupancies
\begin{equation}
\begin{aligned}
P_{3\mathrm{off};1}&=\frac{1}{3N}\sum_{\langle ij\rangle}\langle n_{i2}n_{i3}n_{j1}\rangle\\
P_{3\mathrm{off};3}&=\frac{1}{3N}\sum_{\langle ij\rangle}\langle n_{i1}n_{i2}n_{j3}\rangle
\end{aligned}
\end{equation}
  where $N$ is the number of lattice sites. The dependence of occupancies $P_{3}$, $P_{3\mathrm{off};1}$ and $P_{3\mathrm{off};3}$ on the coupling strength $|U'|$ are presented in Fig.~\ref{occupation}. In the whole range of $3 \leqslant |U^\prime| \leqslant 6$, $P_{3} \gg P_{3\mathrm{off}}>0$, which manifests that minority off-site trions coexist with majority on-site trions.  Evidently, when $|U'|<5.5$, $P_{3\mathrm{off};3} > P_{3\mathrm{off};1}$, which implies that color 3 and colors (1, 2) carried by off-site trions tend to occupy different sublattices (i.e. formation of N\'eel-like off-site trions). The QMC results agree with our physical intuition coming from the first-order perturbation theory of the two-site Hubbard model. Moreover, with the decrease of $|U^\prime|$, $P_{3\mathrm{off}}$ increases while $P_{3}$ decreases, which displays the physical picture that interaction-anisotropy enhanced density fluctuations promote the formation of off-site trions from on-site trions.

We next show that the growing number of N\'eel-like off-site trions caused by interaction anisotropy can develop long-range N\'eel order. Along the same line as Ref. \cite{Wang2014}, the N\'eel configuration in our three-color model is contributed by the net magnetic moments of the off-site trion: the two-fermion end with net colors (1, 2) is in sublattice A, while the one-fermion end with net color 3 is in sublattice B. The N\'eel moment operator on site $i$ is then defined as
\begin{equation}
m_{i}=\frac{1}{4}\left(n_{i1}+n_{i2}-2n_{i3}\right)
\end{equation}
which yields the N\'eel moment $\langle m_{i}\rangle=(-1)^{i}\frac{1}{2}$. The N\'eel order parameter is defined as a sum over the correlations between N\'eel moment operators
\begin{equation}
m_{Q}=\frac{1}{N}\sqrt{\sum_{ij}(-1)^{i+j} \langle m_{i}m_{j} \rangle}.
\end{equation}
To investigate the charge distribution, we also calculate the staggered order parameters \cite{Miyatake2010,Inaba2013,Yanatori2016,Yanatori2016b}
\begin{equation}
\begin{aligned}
\label{staggered order}
M_{1}&=\frac{1}{N}\sqrt{\sum_{ij}(-1)^{i+j}\langle n_{i1}n_{j1}\rangle}\\
M_{3}&=\frac{1}{N}\sqrt{\sum_{ij}(-1)^{i+j}\langle n_{i3}n_{j3}\rangle},
\end{aligned}
\end{equation}
and the trionic CDW order parameter \cite{xu2019,li2021}
\begin{equation}
\label{CDW}
D=\frac{1}{N}\sqrt{\sum_{i,j}(-1)^{i+j}C(i,j)},
\end{equation}
\begin{figure}[t]
  \centering
  \includegraphics[width=0.7\linewidth]{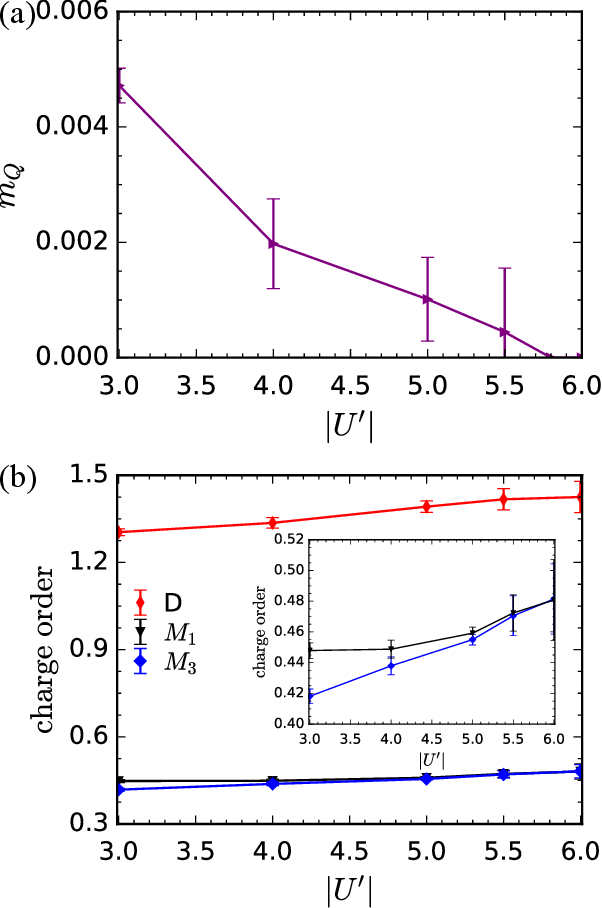}
  \caption{(a) The plot of the N\'eel order parameter $m_{Q}$ as a function of $|U^\prime|$; (b) The plots of trionic CDW order parameter $D$, the staggered order parameters $M_{1}$ and $M_{3}$ as a function of $|U^\prime|$. The inset is a zoom-in view of $M_{1}$ and $M_{3}$ curves.}
  \label{scaling orders}
\end{figure}
where $C(i,j)=\sum_{\alpha,\beta}\langle n_{i\alpha}n_{j\beta}\rangle$ is the density-density correlation function. In the coupling region $3\leqslant |U^\prime| \leqslant 6$,
the N\'eel order parameter $m_{Q}$, trionic CDW order parameter $D$, staggered order parameters $M_{1}$ and $M_{3}$ are obtained by finite-size scaling (See Supplemental Material \cite{supplemental}), and are respectively plotted as a function of $|U'|$ in Fig.~\ref{scaling orders}. As shown in Fig.~\ref{scaling orders} (a), when the coupling strength $|U^\prime|<5.5$, N\'eel ordering develops ($m_{Q} > 0$). In Fig.~\ref{scaling orders} (b), on one hand, in the specified coupling regime trionic CDW order parameter $D$ is slightly smaller than its limiting value, 1.5 (In this limit only on-site trions exist and they occupy one sublattice), indicating that strong trionic CDW order develops in the mixture of on-site trions and off-site trions; on the other hand, when $|U^\prime|<5.5$, staggered order parameters obey $M_{3} < M_{1}$, which indicates that on the background of the trionic CDW order, the one-fermion ends of off-site trions carrying net color 3 tend to be the nearest neighbor of on-site trions and thus occupy a sublattice (Two-fermion ends of off-site trions carrying net colors (1, 2) then occupy the other sublattice), reflecting the N\'eel ordering of off-site trions. When $|U^\prime| \geqslant 5.5 $, $M_1$ and $M_3$ curves coincide perfectly with each other (Fig.~\ref{scaling orders} (b)), vanishing the N\'eel-like configuration of an off-site trion, and accordingly N\'eel order $m_Q=0$ (Fig.~\ref{scaling orders} (a)). Our QMC results clearly show that the control of coupling strength $|U^\prime|$ induces the N\'eel ordering of off-site trions on the background of strong trionic CDW order.

\begin{figure}[t]
  \centering
  \includegraphics[width=0.85\linewidth]{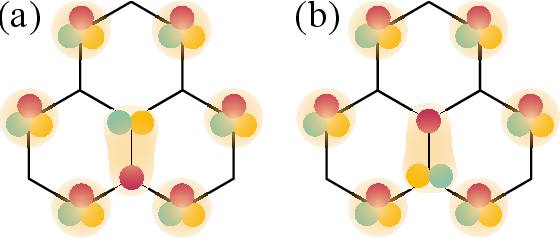}
  \caption{Two possible orientations of an off-site trion on the background of the trionic CDW order: (a) The one-fermion end of the off-site trion is the nearest neighbor to on-site trions; (b) The two-fermion end of the off-site trion is the nearest neighbor to on-site trions.}
  \label{hopping}
\end{figure}

We can intuitively comprehend, from the perspective of energy, the underlying physics of N\'eel ordering of off-site trions. On the background of the trionic CDW order, an off-site trion only has two possible orientations: (1) its one-fermion end is the nearest neighbor to on-site trions, as illustrated in Fig.~\ref{hopping} (a); and (2) its two-fermion end is the nearest neighbor to on-site trions, as depicted in Fig.~\ref{hopping} (b). The Pauli exclusion blocks more hopping channels in Fig.~\ref{hopping} (b) than in Fig.~\ref{hopping} (a). Hence, to lower the total kinetic energy, N\'eel-like off-site trions (resulting from breaking SU(3) symmetry of Hubbard interactions) tend to follow the spatial arrangement outlined in Fig.~\ref{hopping} (a), which is exactly the same as QMC results. Since the two ends of N\'eel-like off-site trions occupy different sublattice, the long-range N\'eel ordering develops when the number of off-site trions continues to rise due to the enhanced density fluctuation via tuning the coupling of color 3 with colors (1, 2) $|U^\prime|$.

\label{pairing}
\textit{CSF order breaking}.--- When the color-3 fermion is weakly coupled to the CSF state of fermions carrying colors (1, 2) (i.e. $|U^\prime| \ll |U| $), DMFT studies \cite{Miyatake2011,Okanami2014} suggest a direct phase transition between the CSF state and the on-site trionic state at finite $|U^\prime|$. We argue that CSF order vanishes at very small $|U^\prime|$ accompanied by the symmetry reduction from SO(4)$\otimes$U(1) to SU(2)$\otimes$U(1), and the on-site trionic phase does not emerge immediately after breaking CSF order since $|U^\prime$| is very small. Our point can be verified by QMC simulations.

We first define the $s$-wave pairing (i.e. CSF) structure factor
\begin{figure}[b]
  \centering
  \includegraphics[width=1\linewidth]{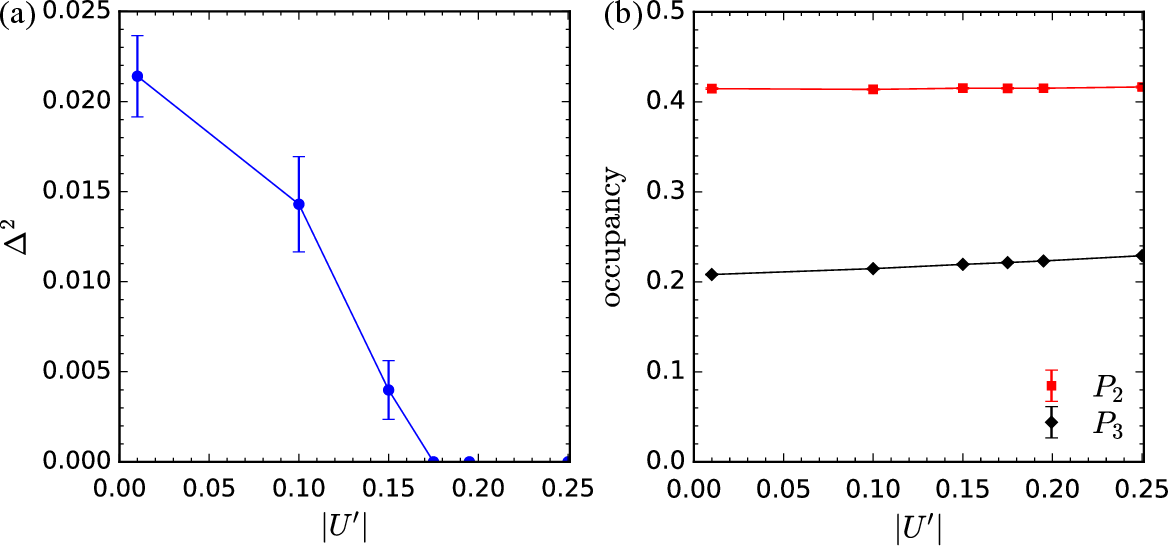}
  \caption{ (a) The square of CSF order parameter $\Delta^2$, (b) the double occupancy $P_{2}$ and the triple occupancy $P_{3}$ are plotted as a function of coupling strength $|U^\prime|$. $P_{2}$ and $P_{3}$ are calculated on the $L=9$ lattice and $\Delta^2$ is obtained by finite-size scaling (See Supplemental Material \cite{supplemental}). Error bars are smaller than the data points in (b). }
  \label{phase coherance}
\end{figure}
\begin{equation}
\Delta_{\mathrm{s}}=\frac{1}{N}\sum_{ij}\langle c_{i1}^{\dagger}c_{i2}^{\dagger}c_{j2}c_{j1}+h.c.\rangle.
\end{equation}
The CSF order parameter is then expressed as $\Delta=\sqrt {\Delta_{s}/N}$.
To characterize the densities of pairs and on-site trions, we also calculate the double occupancy
\begin{equation}
P_{2}=\frac{1}{N}\sum_{i}\langle n_{i1}n_{i2}\rangle,
\end{equation}
and the triple occupancy $P_{3}$ for various small $|U^\prime|$ around the SO($4$)$\otimes$U($1$) symmetric point. The square of CSF order parameter $\Delta^2$, the double occupancy $P_{2}$ and the triple occupancy $P_{3}$ are plotted as a function of coupling strength $|U^\prime|$ in Fig.~\ref{phase coherance}.
When the coupling $|U^\prime|\geqslant 0.175$, the CSF order vanishes while the double occupancy $P_{2}$ is almost unchanged near the transition point and still much larger than the triple occupancy $P_{3}$ (which is almost unchanged as well). This evidently manifests that the dominant entities are pairs rather than on-site trions immediately after the vanishing of CSF order, which disagrees with the DMFT results \cite{Miyatake2011,Okanami2014}.

\begin{figure}[t]
  \centering
  \includegraphics[width=0.75\linewidth]{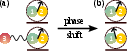}
  \caption{The phase coherent pairs carrying colors (1, 2) (a) are scattered by the color-3 fermions, leading to incoherent doubleons (b).}
  \label{dephasing}
\end{figure}

The mechanism of vanishing CSF order around the SO($4$)$\otimes$U($1$) symmetric point can be intuitively understood in virtue of the concept of Bose-Einstein condensate (BEC) of pairs \cite{Lee2009,Fontenele2022}. In the subsystem of fermions carrying colors (1, 2), at strong attractive couplings, pairs develop into the phase coherent BEC state
\begin{equation}
|\psi\rangle=e^{e^{i\phi}\sum_{i} c_{i1}^{\dagger}c_{i2}^{\dagger}}|0\rangle,
\end{equation}
in which the average of the pairing structure factor is $\langle\psi|\Delta_{\mathrm{s}}|\psi\rangle=4$. However, when a pair carrying colors (1, 2) is scattered by a color-3 fermion on a site, as is illustrated in Fig.~\ref{dephasing} (a), the phase on the scattering site is shifted, resulting that the BEC of pairs evolves into a phase incoherent state (doubleons)
\begin{equation}
|\psi^{\prime}\rangle=e^{\left(e^{i\phi} c_{11}^{\dagger}c_{12}^{\dagger}+e^{i(\phi+\Delta \phi)} c_{21}^{\dagger}c_{22}^{\dagger}\right)}|0\rangle,
\end{equation}
as is illustrated in Fig.~\ref{dephasing} (b). In the phase incoherent state, the pairing structure factor decreases since $\langle\psi^{\prime}|\Delta_{\mathrm{s}}|\psi^{\prime}\rangle=2+2\cos(\Delta \phi)< 4$. This microscopic picture of the CSF-breaking mechanism illustrates that, before the development of trionic phase, the CSF order vanishes along with the emergence of incoherent doubleons, because the phase shift caused by the color-3 fermion scattering destroys the phase coherence in the BEC of pairs, leading to suppression of the pairing correlation.

\textit{Summary}.--- We have performed the sign-problem-free determinant QMC simulations at $T=0.1t$ to investigate the phase transitions in the half-filled attractive three-color Hubbard model on a honeycomb lattice, where Hubbard interactions are color-dependent and set as $|U_{1,2}|=|U|=6$ and $|U_{1,3}|=|U_{2,3}|=|U^\prime|$. The coupling $|U^\prime|$ serves as a control parameter inducing phase transitions. In the coupling regime $3 \leqslant |U^\prime| \leqslant 6$ where on-site and off-site trions coexist, our QMC simulations demonstrate coexisting N\'eel and CDW orders driven by interaction anisotropy $|U^\prime| < |U|$. The N\'eel order reflects the long-range correlation of N\'eel-like off-site trions, while CDW order is modulated by both on-site and off-site trions. In the weak coupling regime $|U^\prime| < 0.25$, our QMC result shows that double-occupancy probability is almost $|U^\prime|$-independent and much greater than the triple-occupancy probability, which implies that the vanishing of the CSF order is not immediately accompanied by the emergence of the on-site trionic phase.

The coexistence of spin/charge orders has long been a research focus in condensed matter physics, but yet absent in cold atom physics. We realize the coexisting N\'eel and CDW orders in the half-filled attractive three-color Hubbard model via tuning the coupling $|U^\prime|$ ($<|U|$) at $|U|=6$. The system of attractive three-color fermions can accommodate multiple Fermi species (unbound single fermion, on-site trion and off-site trion) along with the variation of couplings, which makes it unique from other correlated fermionic systems.  Our work opens up a new avenue for exploring many-body physics of attractive three-color fermions depending on the interplay of three Fermi species, which may host exotic quantum phases and phase transitions.

\acknowledgments
This work is financially supported by the National
Natural Science Foundation of China under Grants No. 11874292, No. 11729402, and No. 11574238.
We acknowledge the support of the Supercomputing Center of Wuhan University.

%
\end{document}